\documentclass{article}
\usepackage{spconf,amsmath,graphicx}
\usepackage{bm}
\usepackage{amsthm,amssymb,amsfonts}

\usepackage{amsmath}
\usepackage{diagbox}
\usepackage{graphicx}
\usepackage{float}
\newfloat{figtab}{htb}{fgtb}
\makeatletter
\newcommand\figcaption{\def\@captype{figure}\caption}
\newcommand\tabcaption{\def\@captype{table}\caption}
\makeatother

\usepackage{amssymb,amsthm}
\usepackage{color}
\usepackage{graphicx}
\usepackage{empheq}
\usepackage[linesnumbered,ruled,lined]{algorithm2e}
\usepackage[outdir=./]{epstopdf}
\usepackage{caption}
\usepackage{lipsum}
\usepackage{cite}
\usepackage{multirow}
\usepackage{subcaption}
\usepackage{fancyhdr}

\newtheorem{theorem}{Theorem}

\newtheorem{lemma}{Lemma}

\SetKwInput{Input}{input}
\SetKwInput{Output}{output}

\usepackage{array}
\newcolumntype{L}[1]{>{\raggedright\let\newline\\\arraybackslash\hspace{0pt}}m{#1}}
\newcolumntype{C}[1]{>{\centering\let\newline\\\arraybackslash\hspace{0pt}}m{#1}}
\newcolumntype{R}[1]{>{\raggedleft\let\newline\\\arraybackslash\hspace{0pt}}m{#1}}

\usepackage{amsmath,amsfonts}

\usepackage{array}
\usepackage{textcomp}
\usepackage{stfloats}
\usepackage{url}
\usepackage{verbatim}
\usepackage{graphicx}
\usepackage{stfloats}
\usepackage{balance}
\usepackage{booktabs}
\usepackage{etoolbox}
\makeatletter\patchcmd{\@makecaption}{\scshape}{}{}{}

\usepackage{amsthm,amssymb,amsfonts}
\usepackage{bm}
\usepackage{cases}
\def\x{{\mathbf x}}

\title{OPTIMAL BEAMFORMING STRUCTURE FOR RATE SPLITTING MULTIPLE ACCESS}
\name{Tianyu Fang$ ^{\dagger} $ , Yijie Mao$ ^\dagger $\thanks{This work has been supported in part by the National Nature Science Foundation of China under Grant 62201347; and in part by Shanghai Sailing Program under Grant 22YF1428400.\textit{(Corresponding author: Yijie Mao.)}}}
\address{	$ ^\dagger $School of Information Science and Technology, ShanghaiTech University, Shanghai, China\\}
\begin{document}
%
\maketitle
\begin{abstract}
In this paper, we aim at maximizing the weighted sum-rate (WSR) of rate splitting multiple access (RSMA) in multi-user multi-antenna transmission networks through the joint optimization of rate allocation and beamforming. Unlike conventional methods like weighted minimum mean square error (WMMSE) and standard fractional programming (FP), which tackle the non-convex WSR problem iteratively using disciplined convex subproblems and optimization toolboxes, our work pioneers a novel toolbox-free approach. For the first time, we identify the optimal beamforming structure and common rate allocation for WSR maximization in RSMA by leveraging FP and Lagrangian duality. Then we propose an algorithm based on FP and fixed point iteration to optimize the beamforming and common rate allocation without the need for optimization toolboxes. Our numerical results demonstrate that the proposed algorithm attains the same performance as standard FP and classical WMMSE methods while significantly reducing computational time. 
\end{abstract}
\thispagestyle{empty}
\vspace{-0.4cm}
\section{Introduction}
\label{sec:intro}
\vspace{-0.2cm}
 Rate splitting multiple access (RSMA) has recently gained prominence as an influential interference management and non-orthogonal transmission technique in the next generation wireless network \cite{bruno2022tutorial,Dizdar2020}. This scheme involves splitting user messages into common and private sub-messages at the transmitter and allowing users to decode the common streams of interfering users as well as the intended private streams. RSMA allows partial decoding of the interference and partial treatment of the remaining interference as noise \cite{Clerckx2016}. The presence of an additional common stream in 1-layer RSMA presents two challenges for the weighted sum-rate (WSR) maximization. First, there is no closed-form expression available for the corresponding common rate. Second, the allocation of this common rate to users creates a coupled common rate allocation problem within the WSR problem.  
 
 Joint resource optimization for beamforming design and common rate allocation has been widely studied in existing works of RSMA to maximize the WSR. Globally optimal resource optimization algorithms \cite{Matth2022Globally,Wang2023} demonstrate appealing performance, but their limitation lies in their exponential computational complexity. Consequently, the state-of-the-art works primarily concentrate on designing suboptimal algorithms that can attain nearly optimal performance. These algorithms draw inspiration from various techniques, such as the weighted minimum mean square error (WMMSE) approach \cite{Hamdi2016,mao2018,Mao2019uni-multicast}, successive convex approximation (SCA), \cite{Mao2019uni-multicast,Mao2020,Li2020}, semidefinite relaxation (SDR) \cite{Fu2020,Dizdar2020a}, and the alternating direction method of multipliers (ADMM) \cite{Onur2021Radar,Xu2021,CernaLoli2021}. However, their practical applicability remains limited due to the computational complexity associated with iterative use of CVX optimization solvers \cite{grant2014cvx}.
 
  While existing methods can lead to workable solutions, they fall short in providing a comprehensive understanding underlying the solution structure and often grapple with huge computational demands. One effective approach to develop an algorithm that balances computational efficiency with nearly optimal performance is to discover and leverage the inherent optimal beamforming structure and common rate allocation for the WSR problem. However, this has not yet been achieved for RSMA. To fill this gap, we consider the WSR maximization problem of 1-layer RSMA for multi-antenna broadcast channel (BC). The main contributions of this work are twofold: 1) This is the first work that characterizes the optimal beamforming structure and common rate allocation for the WSR problem of RSMA. 2) By exploiting such optimal solution structure, we propose an efficient CVX-free algorithm to solve the WSR problem of RSMA. Simulation results demonstrate that the proposed algorithm attains the same performance as traditional FP and classical WMMSE methods while significantly reducing computational time. 
\vspace{-0.4cm}
\section{System Model and Problem Formulation}
\label{sec:format}
\vspace{-0.2cm}
In this study, we focus on the widely adopted RSMA strategy known as 1-layer RSMA \cite{Mao2022} in a multiple-input single-output (MISO) BC, comprising a base station (BS) and $ K $ users, which are indexed by $ \mathcal K=\{1,2,\cdots,K\} $. The BS has $ L $ antennas while each user possesses a single-antenna. Each user message $ W_k $ is split into a common sub-message $ W_{c,k} $ and a private sub-message $ W_{p,k} $. The common sub-messages are combined and encoded into a single common stream $ s_0 $ while the private sub-messages are independently encoded into the private streams $ \{s_1,\cdots,s_K\} $. It is assumed that each stream in $ \{s_0,s_1,\cdots, s_K\} $ satisfies $\mathbb{E}[s_ks_k^H]=1$, and $ \mathbb{E}[s_ks_i^H]=0, \forall k\neq i$,  $k,i\in\mathcal{K}\cup\{0\}$. Denote the corresponding beamforming vector of $ s_k $ as $ \mathbf w_k\in\mathbb{C}^{L} $, the signal transmitted by the BS can be expressed as $ \mathbf{x}=\sum_{k=0}^{K}\mathbf{w}_ks_k $. The transmitted signal $ \x $ should satisfy the transmit power constraint $ \sum_{k=0}^K \|\mathbf w_k\|^2\leq P_t $, where $ P_t $ represents the upper limit of the transmit power as the BS.

The received signal at each user $ k $ is given by $ 	\label{signal}
	y_k=\mathbf h_k^H\mathbf x+n_k,  k\in\mathcal{K} $, where $ \mathbf{h}_k^H\in\mathbb{C}^{1\times L} $ refer to the downlink channel vector from the BS to user $ k $ and $ n_k\sim \mathcal{CN}(0,\sigma_k^2) $ denotes the additive white Gaussian noise (AWGN) with $ \sigma_k^2 $ being the noise power at user $k$. Prior to decoding their intend private stream $ s_k $, each user first decodes the common stream $ s_0 $ and employs the successive interference cancellation (SIC) technique to eliminate the interference from common stream. The received signal-to-interference-plus-noise ratio (SINR) of the common stream and private streams at user $ k $ are given as
 \begin{equation}
 	\label{eq:SINRs}
  	\gamma_{i,k}=\frac{|\mathbf h_k^H\mathbf w_i|^2}{\sum_{j=1,j\neq i}^K|\mathbf h_k^H\mathbf w_j|^2+\sigma_k^2},\,\,\, k\in\mathcal K,i\in\{0,k\},
 \end{equation}
where the subscript $\{j=1,j\neq i\}$ reduces to  $\{j=1\}$ when $i=0$ or reduces to $ \{j=1,j\neq k\} $ when $ i=k, k\in\mathcal K $. The corresponding achievable rates of common and private streams are given as  $ r_{i,k}=\log\left(1+\gamma_{i,k}   \right), k\in\mathcal{K},i\in\{0,k\}$. Since the common stream must be decoded by all users, the common rate is less than or be equal to the worst-case achievable rate in $ \{r_{0,1},\cdots,r_{0,K}\} $. For user $ k $, only a portion of common stream is required, and we denote the rate for this portion as $ c_k $, also referred to as the fraction of the common rate allocated to user $ k $. The sum of these fractions should not exceed the common rate, i.e., $ \sum_{i=1}^{K}c_i\leq r_{0,k},\forall k\in\mathcal{K}$. Hence, the overall transmission rate for user $ k $ is the combination of the rate allocated to the common stream and the private rate obtained from decoding their private streams, expressed as $r_k^{\text{tot}} = c_k + r_{k,k}$.
\par Let $\delta_k  $ denote to the weight of user $ k $, the WSR maximization problem for 1-layer RSMA is formulated as \cite{Hamdi2016}:
\vspace{-0.2cm}
\begin{subequations}
	\label{P1}
	\begin{align}
		\max\limits_{\mathbf W,\mathbf c}\,\, &\sum\limits_{k=1}^K \delta_k(c_k+r_{k,k})\\
		\text{s.t.}\,\,		\label{leq: allocation constraint} &c_k\geq 0,\,\,\, \forall k\in \mathcal{K},\\
		\label{leq: common rate}	&\sum_{i=1}^K c_i\leq r_{0,k},\,\,\, \forall k\in\mathcal{K},\\
		\label{eq:power constraint}	& \mathrm{tr}(\mathbf {WW}^H)\leq P_{t},
	\end{align}
\end{subequations}
where $\mathbf W=[\mathbf w_0,\mathbf w_1,\cdots,\mathbf w_K]$ and $ \mathbf c=[c_1,c_2,\cdots,c_K]^T $. Constraint (\ref{leq: allocation constraint}) shows the rate allocated to user $ k $ should be non-negative and constraint (\ref{leq: common rate}) refers to the common rate constraint. The transmit power constraint at the BS is represented in (\ref{eq:power constraint}).
\section{Optimal beamforming structure for RSMA}
\label{sec:pagestyle}

In this section, We focus on obtaining the optimal common rate allocation and beamforming structure for problem \eqref{P1}. 
\vspace{-0.4cm}
\subsection{Optimal Common Rate Allocation}
\vspace{-0.2cm}
\par  To obtain the optimal common rate allocation, we discover the following Proposition 1.

\textbf{Proposition 1.}
	\textit{The optimal common rate allocation $ \mathbf c^{\circ} $ for the WSR maximization problem of RSMA in \eqref{P1} is given by 
	\begin{equation}
		\label{opt a}	c_i^{\circ}=\left\{	\begin{aligned}
			&\min_{k\in\mathcal K}\{r_{0,k}\},\,\,\, \text{if}\,\,\, i=\arg\max_{i\in\mathcal K}\{\delta_i\},\\
			&0,\quad \quad\quad\quad\quad\!\!\!\!\!\!\,\,\,\text{otherwise.}
		\end{aligned}\right.	\vspace{-0.3cm}
	\end{equation}}
\par \textit{Proof:} For any given beamforming matrix $ \mathbf W $, problem (\ref{P1}) can be simplified to a straightforward linearly programming. By considering $ r_{k,k} $ and $ r_{0,k} $ as constants in \eqref{P1}, it is easy to obtain that
\vspace{-0.2cm}
\begin{equation}\label{linear programing}
	\sum_{k=1}^K \delta_k c_k \leq \max_{k\in\mathcal K}\{\delta_k\}\cdot\sum_{k=1}^K c_k\leq \max_{k\in\mathcal K}\{\delta_k\}\cdot\min_{k\in\mathcal K}\{r_{0,k}\}.
\end{equation}
The equalities in \eqref{linear programing} are achieved if and only if $\mathbf c=\mathbf c^{\circ}  $, which completes the proof.  $ \hfill\blacksquare $ 

Note that if there are users with the same highest weight, then any arbitrary allocation of the common rate among these users would yield the same WSR, and our proposed solution is one of the optimal common rate allocation schemes. Therefore, there is no need to delve into various special cases of user weights.

\par By substituting \eqref{opt a} back into problem \eqref{P1}, we obtain the following equivalent problem:
	\vspace{-0.2cm}
\begin{subequations}
	\vspace{-0.3cm}
	\label{P pureW}
	\begin{align}
		\max\limits_{\mathbf W,y}\,\, &\max_{k\in\mathcal{K}} \{\delta_k\} \cdot y+ \sum\limits_{k=1}^K \delta_kr_{k,k}\\
		\text{s.t.}\,\,		
		&y\leq r_{0,k},\,\,\, \forall k\in\mathcal{K},\\
		\label{P Wc2}	& \mathrm{tr}(\mathbf {WW}^H)\leq P_{t},
	\end{align}
\end{subequations}
where $ y $ is a slack variable to represent $ \min_{k\in\mathcal K}\{ r_{0,k}\} $. In the following, we focus only on the beamforming optimization since the common rate allocation has been removed from problem \eqref{P pureW} according to Proposition 1. 
\vspace{-0.4cm}
\subsection{Fractional Programming Method}\label{FPM}
\vspace{-0.2cm}
\par We first apply the FP technique named Lagrangian dual and quadratic transforms proposed in \cite{Shen2018} to transform problem \eqref{P pureW} into a more tractable equivalent form. By introducing the auxiliary variable  $ \alpha_{i,k} $,  one can reformulate the  achievable rate $ r_{i,k} $ at user $ k $ for decoding $ s_i, i\in\{0,k\}$ as
\begin{equation}
	\label{eq:FP1}
	f_{i,k}(\mathbf W,\alpha_{i,k})\triangleq \log(1+\alpha_{i,k})-\alpha_{i,k}+\frac{(1+\alpha_{i,k})\gamma_{i,k}}{1+\gamma_{i,k}}.
\end{equation}
Furthermore, through introducing auxiliary variable  $ \beta_{i,k} $ to decouple the fractional term in (\ref{eq:FP1}), the recast achievable rate of user $ k $ to decode stream $ s_i, i\in\{0,k\}$ is obtained as
\begin{equation}
	\label{eq:FP2}
	\begin{split}
	&g_{i,k}(\mathbf W,\alpha_{i,k},\beta_{i,k})	\triangleq\!2\sqrt{1+\alpha_{i,k}}\Re\{\beta_{i,k}^H\mathbf h_k^H\mathbf w_i  \}\!-\alpha_{i,k}\\
	&-\!|\beta_{i,k}|^2\left(\sum_{j\in\mathcal{K}\cup\{i\}}\!\!\!|\mathbf h_k^H\mathbf w_j|^2+\sigma_k^2\!\!\right)+\log(1+\alpha_{i,k}).
	\end{split}
\end{equation}
Replacing $ r_{i,k} $ in (\ref{P pureW}) with $ g_{i,k}(\mathbf W,\alpha_{i,k},\beta_{i,k})$ leads to the following equivalent problem:
	\begin{subequations}
		\label{P3}
		\vspace{-0.3cm}
		\begin{align}
			\max\limits_{\mathbf W,y,\bm{\alpha}, \bm{\beta}}\,\, &\max_{k\in\mathcal K} \{\delta_k\}\cdot y+\sum\limits_{k=1}^K \delta_kg_{k,k}(\mathbf W,\alpha_{k,k},\beta_{k,k})\\
			\label{P3c1}	\text{s.t.}\,\,	&y\leq g_{0,k}(\mathbf W,\alpha_{0,k},\beta_{0,k}),\,\,\, \forall k\in\mathcal{K},\\
			\label{P3c3}	& \mathrm{tr}(\mathbf {WW}^H)\leq P_{t},
		\end{align}
	\end{subequations}
	where $\bm{\alpha}\triangleq\{\alpha_{i,k}|i\in\{0,k\}, k\in\mathcal{K}\}$ and $\bm{\beta}\triangleq\{\beta_{i,k}|i\in\{0,k\}, k\in\mathcal{K}\}$. For more details about the FP method, readers can refer to \cite{Shen2018}.

\par Problem \eqref{P3} exhibits block-wise convexity, with the optimization of both the beamforming matrix and the introduced auxiliary variables performed within an alternative optimization (AO) framework. Specifically, by setting $ \partial f_{i,k}/\partial \alpha_{i,k} $ and $ \partial g_{i,k}/\partial \beta_{i,k} $ to zeros respectively, the optimal $ \alpha_{i,k} $ and $ \beta_{i,k} $ can be obtained by 
	\vspace{-0.3cm}
\begin{equation}
	\label{eq:optalpha}
	\vspace{-0.2cm}
	\begin{aligned}
		\alpha_{i,k}^{\star}&=\gamma_{i,k},\\
		\beta_{i,k}^{\star}&=\frac{\sqrt{1+\alpha_{i,k}}\mathbf{h}_{k}^{H}\mathbf w_i}{\sum_{j\in\mathcal{K}\cup\{i\}}|\mathbf h_k^{H}\mathbf w_j|^2+\sigma_k^2}. 
	\end{aligned}
\end{equation}
For given $ \{ \bm\alpha,\bm\beta\} $, the optimal $ \{\mathbf W,y\}$ can be obtained by solving the following convex problem:
\begin{subequations}
	\label{P5} 
	\vspace{-0.2cm}
	\begin{align}
		\max\limits_{\mathbf W,y}\,\, &\max_{k\in\mathcal K} \{\delta_k\}\cdot y+\sum\limits_{k=1}^K \delta_kg_{k,k}(\mathbf W)\\
		\text{s.t.}\,\,&	y\leq g_{0,k}(\mathbf W),\,\,\, \forall k\in\mathcal{K} \label{P5:C1}\\
		& \mathrm{tr}(\mathbf W\mathbf W^H)\leq P_t \label{P5:C2}.
	\end{align}
\end{subequations}
The approach of using AO to solve problem (\ref{P3}) and using the standard solvers in CVX to solve problem \eqref{P5} is known as the standard FP method and summarized as follows:
\vspace{-0.2cm}
\begin{enumerate}
	\item  Set initial feasible $ \mathbf W^{[0]} $; $ n= 1$;\vspace{-0.2cm}
	\item Update the auxiliary variables $ \alpha_{i,k}^{[n]} $ and $ \beta_{i,k}^{[n]} $ by \eqref{eq:optalpha} for $ k\in\mathcal K,i=\{0,k\} $;\vspace{-0.2cm}
	\item Update $ \mathbf W^{[n]} $ and $ y^{[n]} $ by solving problem \eqref{P5} optimally with CVX;\vspace{-0.2cm}
	\item Set $ n=n+1 $; Return to step 2 until convergence.\vspace{-0.2cm}
\end{enumerate}
The above standard FP approach is ensured to converge to a stationary (locally optimal) solution  of the original problem (\ref{P1}) \cite{Shen2018}.
\subsection{Optimal Beamforming Structure}
\label{subsection OBS}

\par  In this subsection, we leverage Lagrange duality and Karush-Kuhn-Tucker (KKT) conditions to derive the optimal beamforming structure for problem \eqref{P5} and subsequently for problem \eqref{P1}. To the best of our knowledge, this paper is the first to uncover the optimal beamforming structure for RSMA. By introducing Lagrangian dual variables $\bm\lambda=[\lambda_1,\cdots,\lambda_K]^T $ and $\mu$ respectively  for (\ref{P5:C1}) and (\ref{P5:C2}), the Lagrange function for problem (\ref{P5}) is defined as
\vspace{-0.4cm}
\begin{equation*}
		\begin{split}
			\mathcal{L}(\mathbf W,&y,\bm\lambda,\mu)\triangleq\max_{k\in\mathcal K} \{\delta_k\}\cdot y+\sum\limits_{k=1}^K\delta_k g_{k,k}(\mathbf W)\\
			&-\sum\limits_{k=1}^K\lambda_k(y-g_{0,k}(\mathbf W))-\mu(\mathrm{tr}(\mathbf W\mathbf W^H)-P_t).
		\end{split}
\end{equation*}
 Then, the KKT conditions for problem (\ref{P5}) are given by
\begin{subequations}
	\label{kkt}   
	\begin{align}
		\label{partial W0} \left(\sum_{j=1}^{K}\theta_{0,j}\mathbf h_j  \mathbf h_j^H\!+\mu\mathbf{I}\right )\mathbf w_0&=\sum_{j=1}^{K}d_{0,j} \mathbf h_j,\\
		\label{partial Wk}	\left(\sum_{j=1}^{K}\theta_{j,j}\mathbf h_j  \mathbf h_j^H+\mu\mathbf I\right)\mathbf w_k&=d_{k,k}\mathbf h_k, \forall k\in\mathcal{K},\\
		\label{partial a} \max_{k\in\mathcal K} \{\delta_k\}&=\sum_{k=1}^K\lambda_k\\
		\label{kkt for lambda1}\lambda_k\left(y-g_{0,k}(\mathbf W)\right)&=0,\,\,\,\,\forall k\in\mathcal{K},\\
		\label{kkt for mu}	\mu(\mathrm{tr}(\mathbf W\mathbf W^H)-P_t)&=0,
	\end{align}
\end{subequations}
where $ d_{0,j}, \theta_{0,j},\theta_{j,j},\forall j\in\mathcal K$ and $  d_{k,k} ,\forall k\in\mathcal K$
are respectively defined as
\begin{equation*}\label{key}
	\begin{aligned}
		d_{0,j}&\triangleq\sqrt{1+\alpha_{0,j}}\beta_{0,j}\lambda_j,\!\!\!& \theta_{0,j}&\triangleq\lambda_j |\beta_{0,j}|^2,\\
		d_{k,k}&\triangleq\sqrt{1+\alpha_{k,k}}\beta_{k,k}\delta_k,\!\!\!& \theta_{j,j}&\triangleq\delta_j|\beta_{j,j}|^2+\lambda_j |\beta_{0,j}|^2.
	\end{aligned}
\end{equation*}

\par  The optimal solution for $ y $ is given by $ y^\star= \min_{k\in\mathcal K}\{ g_{0,k}\} $ according to \eqref{P5:C1}. From \eqref{partial W0} and \eqref{partial Wk}, we obtain the closed-form optimal beamforming solution for problem \eqref{P5}, as delineated in Lemma \ref{Pro: Optimal beamforming of subproblem}.
\vspace{-0.2cm}
\begin{lemma}\label{Pro: Optimal beamforming of subproblem}
	The optimal beamforming solution for problem \eqref{P5} is given by
	\vspace{-0.2cm}
	\begin{subequations}
		\label{up W}
		\begin{align}
			&\mathbf w_0^{\star}\!=\left(\mathbf H\bm\Theta_c(\bm\lambda^\star) \mathbf H^H +\mu^\star \mathbf I\right)^{-1}\mathbf H\mathbf d_c(\bm\lambda^\star),\\
			&\mathbf w_k^{\star} \!=d_{k,k}\left(\mathbf H\bm\Theta_p(\bm\lambda^\star) \mathbf H^H +\mu^\star\mathbf I\right)^{-1} \mathbf h_k, k\in\mathcal K,
		\end{align}
	\end{subequations}
	where $ \mathbf H=[\mathbf h_1,\cdots,\mathbf h_K] $ refers to the channel matrix of all users, $\mu^\star$ and $ \bm\lambda^\star $ are the optimal dual variables for problem \eqref{P5}, $\bm\Theta_c(\bm\lambda^\star)$ and $ \bm\Theta_p(\bm\lambda^\star)$ are diagonal matrices with $ \{\theta_{0,1}^\star,\cdots,\theta_{0,K}^\star\} $ and $ \{\theta_{1,1}^\star,\cdots,\theta_{K,K}^\star\} $  respectively along the main diagonals and $ \mathbf d_c(\bm\lambda^\star)=[d_{0,1}^\star,d_{0,2}^\star,\cdots,d_{0,K}^\star]^T $.
\end{lemma}

\par As the global optimal solution lies within the set of locally optimal solutions \cite{Dong2020}, the FP algorithm would converge to the global optimal solution $ \mathbf W^\circ $ of problem (\ref{P1}) if the initial point $  \mathbf W^{[0]} $ is in the vicinity of $ \mathbf W^\circ $. Denote that $ \alpha_{i,k}^\circ=\alpha_{i,k}^\star(\mathbf W^\circ) $ and $ \beta_{i,k}^\circ=\beta_{i,k}^\star(\mathbf W^\circ) $, we establish the following
Theorem \ref{Theorem global optimal}.
\begin{theorem} \label{Theorem global optimal}
	The optimal beamforming structure for the WSR maximization problem of RSMA in \eqref{P1} is given as
	\vspace{-0.1cm}
	\begin{equation}
		\label{global optimal}
		\vspace{-0.1cm}
		\begin{aligned} 
			\mathbf w_c^\circ&=(\mathbf H\bm\Theta_c^\circ \mathbf H^H +\mu^\star\mathbf I)^{-1}\mathbf H \mathbf  d_c^\circ,\\
			\mathbf W_p^\circ&=(\mathbf H\bm\Theta_p^\circ\mathbf H^H +\mu^\star\mathbf I)^{-1}\mathbf H\cdot \mathrm{diag}\{\mathbf d_p^\circ \},
		\end{aligned}
	\end{equation}  
	where $ \bm\lambda^\star,\mu^\star $ are the optimal dual variables and $ \mathbf w_c^\circ\triangleq\mathbf w_0^\circ,\mathbf W_p^\circ\triangleq[\mathbf w_1^\circ,\cdots,\mathbf w_K^\circ], \alpha_{i,k}^\circ=\alpha_{i,k}^\star(\mathbf W^\circ),\beta_{i,k}^\circ=\beta_{i,k}^\star(\mathbf W^\circ)$ , $  \bm\Theta_c^\circ=\bm\Theta_c(\bm\lambda^\star,\bm\beta^\circ), \bm\Theta_p^\circ=\bm\Theta_p(\bm\lambda^\star,\bm\beta^\circ),\mathbf d_c^\circ=\mathbf d_c(\bm\lambda^\star,\bm\alpha^\circ,\bm\beta^\circ)$ and $ \mathbf d_p^\circ=\mathbf d_p (\bm\alpha^\circ,\bm\beta^\circ)  $.
\end{theorem}
\textit{Proof:} The FP-based algorithm would reach to the global optimal solution $ \mathbf W^\circ $ when the initial point $ \mathbf W^{[0]} $ is carefully chosen. In this situation, substituting $ \mathbf W^\circ $ into \eqref{eq:optalpha}, the optimal beamforming of problem \eqref{P5} must be the same optimal beamforming of global solution $ \mathbf W^\circ $, which suggests that the optimal beamforming solutions for problem \eqref{P1} shares the same structure with the optimal solution for subproblem \eqref{P5}.$ \hfill\blacksquare $

\section{Proposed Numerical Algorithms}
\label{sec:typestyle}
\par  In this section, we present a numerical algorithm for computing $\bm\lambda^\star $ and $ \mu^\star $ embedded in \eqref{up W}.  Fixed point iteration (FPI) is a classical method to find the Lagrangian dual variables that satisfy the KKT conditions \cite{Pham2019}. However, the considered WSR problem of RSMA presents an additional equation \eqref{partial a}, which hinders the extension of existing algorithms for determining the optimal values of $ \bm\lambda^\star $ and $ \mu^\star $. To address this issue, we propose an iterative algorithm named hyperplane FPI (HFPI) to find the optimal $ \bm\lambda^\star $ and $ \mu^\star $ based on the \eqref{partial a}, \eqref{kkt for lambda1} and \eqref{kkt for mu}.

\par Denote $ h_{0,k}(\bm\lambda,\mu)\triangleq g_{0,k}(\mathbf W^\star(\bm\lambda,\mu)) $ and assume that user $ m $ achieves the worst-case common rate in each iteration, i.e., $y=h_{0,m}(\bm\lambda,\mu)= \min_{k\in \mathcal{K}} \{h_{0,k}(\bm\lambda,\mu)\}$. The index $ m $ may change in each iteration. We propose the following HFPI algorithm, which updates the dual variables $\bm\lambda$ and $ \mu $ at iteration $[t]$ by
\begin{subequations}
	\label{update_lambda}
	\begin{align}
		\lambda_k^{[t]}&\!=\!\frac{h_{0,m}(\bm\lambda^{[t-1]},\mu^{[t-1]})\!+\!\rho}{h_{0,k}(\bm\lambda^{[t-1]},\mu^{[t-1]})\!+\!\rho}\lambda_k^{[t-1]},  k\neq m, \forall k\in\mathcal{K}\\
		\lambda_m^{[t]}&\!=\!\lambda_m^{[t-1]}\!+\!\sum_{k=1}^K\!\left(\!1\!-\!\frac{h_{0,m}(\bm\lambda^{[t-1]},\mu^{[t-1]})+\rho}{h_{0,k}(\bm\lambda^{[t-1]},\mu^{[t-1]})+\rho}\right)\lambda_k^{[t-1]},\\
		\mu^{[t]}&\!=\!\frac{\mathrm{tr}(\mathbf W^{[t]^H}\mathbf W^{[t]})+\rho}{P_t+\rho}\mu^{[t-1]},
	\end{align}
\end{subequations}
where constant $ \rho\geq 0 $ is employed to enhance convergence stability by effectively reducing the step size. The optimal dual variables $ \bm\lambda^\star $ and $ \mu^\star $ can be obtained by iteratively updating $ \bm\lambda $ and $ \mu $ via equation \eqref{update_lambda} as follows:
\vspace{-0.2cm}
\begin{enumerate}
\item  Initialize $ \bm\lambda^{[0]} $ and $ \mu^{[0]} $; $ t= 1$;\vspace{-0.2cm}
\item Compute $ \bm\lambda^{[t]} $ and $ \mu^{[t]} $ with \eqref{update_lambda};\vspace{-0.2cm}
\item Set $ t=t+1 $; Return to Step 1) until convergence.\vspace{-0.2cm}
\end{enumerate}
Substitute the obtained optimal dual variables into \eqref{up W}, we are able to solve problem \eqref{P5} optimally with the proposed HFPI algorithm instead of using CVX in step 3 of FP method.

\begin{figure}[t]
	
\begin{minipage}[b]{1.0\linewidth}
	\centering
	\centerline{\includegraphics[width=8.5cm]{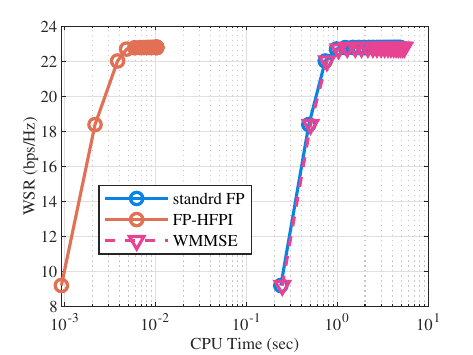}}
\end{minipage}
	\caption{Algorithm performance comparison.}
	\label{fig:res}
\end{figure}

\section{Simulation results}

 In this section, we evaluate the performance of the proposed FP-HFPI algorithm, where the step 3 of standard FP is replaced by the proposed HFPI algorithm. We consider a classic MISO BC with $ L= K=4 $. The channel of user $ k $ is generated i.i.d. as $ \mathbf h_k\sim\mathcal{CN}(\mathbf 0,\mathbf I) $ and the noise variance at user $ k $ is set to $ \sigma_k^2=1 $. The upper transmit power $ P_t $ is $ 100  $W. The constant $ \rho$ is set to $0.5 $ and all  stopping tolerance are set to $ 10^{-4} $. Without loss of generality, we set the user weights as  $ \delta_1=\cdots=\delta_K=1 $. We follow \cite{Hamdi2016} by initializing the private and common beamforming vectors based on maximum ratio transmission (MRT) and the dominant left singular vector of the channel matrix $ \mathbf H $, respectively. All simulation results are averaged over $ 100 $ random channel realizations.
 
 We compare the standard FP, FP-HFPI, and WMMSE in terms of WSR and computation time in Fig.$\,\ref{fig:res} $. The considered WMMSE baseline based on CVX toolbox has been widely used in the existing works of RSMA \cite{Matth2022Globally,Wang2023,Hamdi2016,mao2018,Mao2019uni-multicast,Xu2021}, and it has been shown in reference \cite{Matth2022Globally} that WMMSE achieves almost the same performance compared to the globally optimal algorithm. Fig.$\,\ref{fig:res} $ demonstrate that the proposed FP-HFPI achieves the same WSR performance with the standard FP and WMMSE, while significantly reducing the simulation time by two orders of magnitude.  
\vspace{-0.4cm}
\section{Conclusion}

This paper seeks to reveal the optimal beamforming structure and common rate allocation for the WSR maximization problem in 1-layer RSMA. With the discovered beamforming structure, we further propose an efficient numerical algorithm to determine the involved Lagrangian dual variables. Simulation shows that our proposed method FP-HFPI achieves the same performance as the standard FP ans WMMSE with substantially low complexity.

\newpage

\bibliographystyle{IEEEbib}
\bibliography{Template}

\end{document}